\documentclass[
reprint,
 amsmath,amssymb,
 aps,
floatfix,
]{revtex4-2}
\usepackage{graphicx}
\usepackage{dcolumn}
\usepackage{bm}
\usepackage{amsthm}
\usepackage[utf8]{inputenc}
\usepackage[english]{babel}

\begin{document}

\preprint{APS/123-QED}

\title{Indistinguishable Sub-nanosecond Pulse Generator} 
\author{Pranshu Maan}
\email{Pranshu.Maan@us.bosch.com\\15000 N Haggerty road, Plymouth, Michigan-48170, USA}
\affiliation{Robert Bosch GmbH}
\date{\today}

\begin{abstract}
 Indistinguishable laser pulse is important in realization of qubits in quantum cryptography. Implementation of such cryptography in automotive framework needs to be  compact and cheap. However, their implementation is bulky and expensive. Significant effort has been put into to reduce the form factor of quantum cryptography implementation. We report a compact, low cost, indistinguishable, sub-nanosecond pulse generator with adjusted delay and amplitude, using a Fabry-Perot laser diode. The approach was derived based on algebraic topology formulation of electrical network, and the implementation involves time dependent perturbation of a constant current node, generating tun-able, sub-nanosecond excitation with constant pre-bias. Parameters for simulation model of the laser diode was obtained considering effect of spontaneous emission and relaxation oscillation. The shortest excitation pulse generated was measured to have FWHM of 496ps. Further, indistinguishabilty of two laser pulses are statistically evaluated.\\
 
\end{abstract}

\maketitle
        
\section{Introduction}
Generation of indistinguishable excitation pulses are one of the fundamental requirements in many quantum information related experiments such as: indistinguishable pulses for quantum cryptography application[1][2][3], photon number resolution using APDs[4]. State of art implementations for these experiments are either expensive or too bulky to fit in automotive gateway modules and controllers. This calls for further reduction is form factor and cost.

Different methods have been reported towards implementing quantum cryptography, namely: electro-optic modulators in photonic ICs, driver controlled variable attenuators. In 2015, Wabnig et al.at Nokia labs[9], proposed Quantum Key Distribution(QKD) based scheme using spatial filter to generate indistinguishable pulses, with  Laser diode or LED. The indistinguishable pulses were defined by characteristics of the spatial filter. In 2016, in order to make QKD system compact, Bunandar et al. at MIT[10], proposed photonic integrated chip to realize transmitter and receiver. These photonic ICs comprised of ring resonators.Corresponding to each of these resonators there were modulators that were capable of applying delay to optical pulses. Further, in 2017, Nordholt et al. at Los Alamos National Security[11], proposed another photonic integrated IC based approach. In this, a variable optical attenuator or amplitude modulator was used to reduce average number of photon per pulse. Similarly, in 2017, Yuan et al. at Toshiba[12], created a quantum communication system using variable attenuator to change intensity of the emitted laser pulses.  These approaches often require complex and expensive driving circuits for variable optical attenuators. Direct modulation of laser is another preferred technique to generate indistinguishable excitation[5], where mean number of photon in each pulses can be reduced to less than 1 using a constant optical attenuator, without engendering need of using dedicated driver circuit for the attenuator. However, the state of art was massive, pulse duration obtained was in nanosecond range and pulse characteristics were not adjustable. 

In this paper, inspired by algebraic topology, we show by simulation and experimental measurement, generation of tun-able, indistinguishable, sub-nanosecond excitation. We have achieved shortest injection current FWHM of 496ps using direct modulation of the laser diode. We further demonstrate tun-ability of  excitation pulses to exactly match characteristics in temporal domain.Implementation of quantum cryptography has proven challenging in terms of cost, form factor, and integration into existing infrastructure. This approach can provide low cost and small form factor implementation of quantum cryptography transmitter (and receiver) with greater integration into existing automotive framework.

\section{Theory}
Let us visit some preliminaries pertaining to chain complexes:
\theoremstyle{definition}
\newtheorem{definition}{Definition}[section]
\theoremstyle{remark}
\newtheorem*{remark}{Remark}
\begin {definition}[Boundary operator $\partial$ ]: If $\partial_{n}$ is a boundary operator which is a linear transformation on sequence of vector space $C_{n}$ as
$\partial_{n}$ : $C_{n}\rightarrow C_{n-1}$
, we can write:
\newcommand\myeq{\mathrel{\stackrel{\makebox[0pt]{\mbox{\normalfont\tiny }}}{=}}}
\begin{align*}
C_{n}\xrightarrow{\partial_{n}} C_{n-1} \xrightarrow{\partial_{n-1}} C_{n-2}\ldots \xrightarrow{\partial_{2}} C_{1}\xrightarrow{\partial_{1}} C_{0}\xrightarrow{\partial_{0}}{0}
\end{align*}

as a chain complex if $\partial_{n} \cdot \partial_{n+1}=0.$
If boundary operator on a branch $\omega$ is defined as:
$\partial \omega = \Omega (e-s)$ where $\emph{e}$ is the end node and $\emph{s}$ is the start node of branch $\Omega$, then we can interpret the first operator $\partial_{n}$ on 1-chain complex as nodes. Since node does not have boundary, so second boundary operator on the node, $\partial_{n+1}$, is zero.
\end{definition}

We can now define kernel of a boundary operator.

\theoremstyle{definition}
\begin {definition}
If a k-chain has no boundary, then it is called k-cycles and can be represented as a kernel of the  boundary operator[6]:\\
\begin{equation}
Z_{1}=ker \quad\partial k
\end{equation}

if $k\in Z_{1} $
\begin{equation}
\partial k=0
\end{equation}

\end{definition}
\newtheorem{theorem}{Theorem}[section]
\newtheorem{lemma}[theorem]{Lemma}
\begin{lemma}
Boundary for branches can be expressed as a sum of all the branches that leave and enter a node. 
\end{lemma}
\begin{proof}
\begin{figure}[ht!]
\includegraphics[width=8cm]{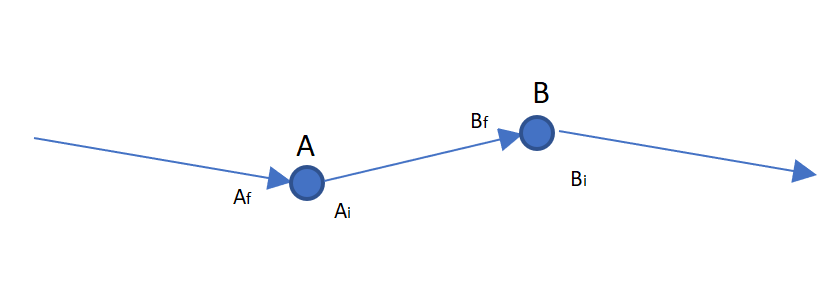}
\caption{\emph{Definition of boundary operator of a branch $\vert A_{f}$ is the branch with A as final node $\vert A_{i}$ is the branch with A as initial node $\vert B_{f}$ is the branch with B as final node. Similarly, $B_{i}$ is the branch with B as initial node.}}
\end{figure}
Fig. 1 represents node A and node B, where $A_{f}$ and $A_{i}$ are branches with node A as final and initial node respectively. Similarly, $B_{f}$ and $B_{i}$ are branches with B as final and initial node respectively.
 If K $\in C_{1}$ is a set of all branches such that
 \begin{equation}
 \partial K=\{k_{a},k_{b} \ldots\}
 \end{equation}
 Here $k_{i}\in C_{0}$ represents branches that enter and leave node i and can be represented as 
 \begin{equation}
 k_{i}=k_{ei}-k_{li}
 \end{equation}
 
 where $k_{ei}$ and $k_{li}$ represent branches that enter and leave node i respectively, irrespective whether it is time-independent or not.
 
 With reference to Fig.1:
 
 \begin{equation}
    \partial K=A_{f}(A-\Delta_{i})+B_{f}(B-A)
 \end{equation}
 where $\Delta_{i}$ represents arbitrary start node of the branch $A_{f}$. Consolidating all the terms pertaining to node A together:
 \begin{equation}
     \partial K=A(A_{f}-B_{f})+B(B_{f})-\Delta_{i}(A_{f})
 \end{equation}
The first term in eqn. 6 pertains to node A, represents branch entering the node (-) branch leaving the node, and it is valid for both constant as well as time dependent branches.
 \end{proof}
Consider an electrical sub-network represented by \{N,B,$\partial$,$R_{dc},R_{perturbation}$\}, where branch B $\in C_{1}$ one chain vector space and node N $\in C_{0}$ zero chain vector space, $\partial : C_{1}\rightarrow C_{0}$ is the boundary map, $R_{dc}$ is the coefficient ring for dc excitation and $R_{perturbation}$ represents coefficient ring for time dependent perturbation on node $\Delta$. If $Z_{1}$ represents kernel of boundary map, then:
 \begin{equation}
    I\in Z_{1}
   \end{equation}
 \begin{equation}
       \partial I =0
   \end{equation}
 \begin{equation}
    \partial I=I_{\delta_{\alpha}}-(I_{\delta_{\beta}}+I_{\delta_{\psi}})=0
   \end{equation}
 \begin{equation}
     I_{\delta_{\psi}}=I_{\delta_{\alpha}}+I_{\delta_{\Delta}}
\end{equation}
where branches that enters or leaves the node $\Delta$, $\delta_{i} \in C_{0},\delta_{\beta}=-\delta_{\Delta}$ is the time dependent negative perturbation applied on node $\beta$ and $\delta_{\alpha}$ is the constant excitation on node $\alpha$.
\begin{figure}[ht!]
\includegraphics[width=8cm]{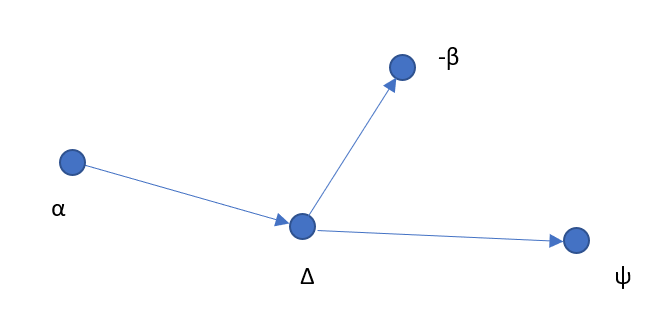}
\caption {\emph{Electrical sub-network for generation of excitation pulse $\vert \Delta$ is the constant current node, $\beta$ is the negative perturbation node, $\alpha$ is the constant excitation node, $\psi$ is the output node $\vert$ Laser diode (or APD for voltage perturbation) is connected at $\psi$ node}}
\end{figure}

Here eqn. 10 can be interpreted as a time dependent perturbation current on node $\psi$ with certain pre-bias current.

In a laser diode,  when injection current is low, spontaneous emission due to carrier relaxation dominates and photon emission varies slowly as a function of injected current. Once injected current approaches lasing threshold, steady state injected current approaches a constant value i.e. spontaneous emission is nearly constant and stimulated emission dominates. This emission in dominant mode is linearly dependent on injection current and each injected carrier results in a photon. Further, relation between resulting injected carrier density and optical intensity can be derived based on laser rate equation for single resonator mode and can be modelled in terms of resistors, capacitors and inductors. Value of these RLC components can be calculated in terms of electron density and photon density[7]. In addition, contribution of spontaneous emission and relaxation oscillation can be also expressed in terms of electrical parameters and electron, photon density[7]. 

\begin{figure}[ht!]
\includegraphics[width=5cm]{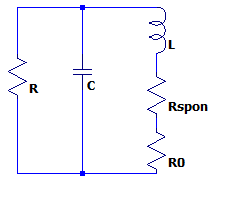}
\caption{\emph{Equivalent circuit model for laser diode[7]$\vert R=2.555\Omega$, L=6.184pH, C=0.3557nF, $R_{se}$=2.811m $\Omega$,$R_{o}$=-5.511m$\Omega$ }}
\end{figure}

\begin{equation}
     R=\dfrac{R_{d}}{n_{photon}+1}
 \end{equation}
\begin{equation}
     L=\dfrac{R_{d}\tau_{photon}}{n_{photon}}
 \end{equation}
 \begin{equation}
     C=\dfrac{\tau_{spon}}{R_{d}}
 \end{equation}
 \begin{equation}
     R_{spon}=\dfrac{\beta R_{d} n_{e}}{n_{photon}^{2}}
 \end{equation}
  \begin{equation}
     R_{o}=\dfrac{-R_{d}\delta}{n_{sat}}\dfrac{1}{[1+{\dfrac{n_{photon}}{n_{sat}}}]^{2}}
 \end{equation}
 
 where $R_{d}= \dfrac{2kT}{q}\dfrac{1}{I_{d}}$ is called differential resistance of the laser diode, $n_{photon}, \tau_{photon}, \tau_{spon}$ are photon density, lifetime of photon and spontaneous emission rate of electrons respectively. $R_{sat}$ represents photon saturation density and $\beta$ represents amount of spontaneous emission that couples into cavity mode of the laser diode. All the circuit components (inductor, resistance due to spontaneous emission and relaxation oscillation) related to optical phenomenon appears in series, as expected from laser rate equation.
Differential resistance decreases with injection current, when below the threshold, and remains nearly constant when injected current is above the threshold.
\section{Simulation}

In this section, we discuss simulation model for analyzing current injection profile into the laser diode. Additionally, using simulation, we show delay and amplitude tuning of the sub-nanosecond perturbation.

Fig. 4 represents simulation model for generating indistinguishable excitation pulse. Shape tuning module comprises of a circuit capable of generating any arbitrary shape excitation. Shape and frequency of the excitation determines frequency and initial characteristics of each perturbation pulse. Shape controlled excitation acts as an input for OPAMP and logic network block.The block comprises of operational amplifier and logical network to implement complimentary output detector through a tun-able differential delay generator.Tun-ability of the block further provides selection of wide range of pulse-width, amplitude and delay for the excitation to generate time-dependent perturbation. Constant node, $\alpha$ from Fig. 2,in the design is excited by a trans-conductance amplifier. Output node $\psi$ is a time dependent perturbation with some pre-bias current as per eqn.10. This excitation node then drives a laser diode. When driving an APD, similar concept can be used to generate tuned voltage excitation.
Design of these blocks are determined by magnitude of current/voltage and timing requirements. OPAMP and logic block was designed to generate perturbation of upto 25mA of peak current. Had this requirement been higher, an additional driver stage would then be required. 

Simulation model for the laser diode was obtained based on discussion in section II. It comprises of a parallel capacitor which arises due to carrier relaxation, differential resistance arising due to carrier injection, and inductance due to photon emission. Additional resistances are due to spontaneous emission and relaxation oscillation in the laser diode (Fig. 3). 
Values for laser diode simulation model were obtained for a laser diode with threshold current value of 18.4mA, at room temperature, and following parameters were considered: $R=2.555\Omega, L=6.184pH, C=0.3557nF, R_{spon}=2.811\Omega,R_{o}=-5.511m\Omega$. 

In addition, a filter stage is designed based on magnitude of attenuation required for filtering of the excitation pulse. Moreover, this stage can also provide feedback for control circuitry in the implementation.

\begin{figure}[ht!]
\includegraphics[width=9cm]{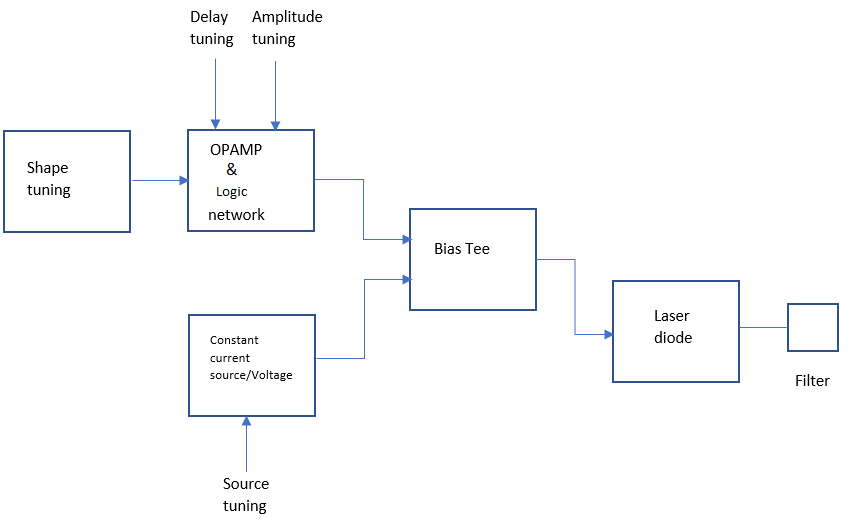}
\caption{\emph{Simulation block diagram $\vert$ model can be adapted for APD instead of a laser diode $\vert$ shape tuning circuit generates square, ramp or any arbitrary shape, OPAMP Logic network is a network comprising of operational amplifiers and FETs (can also be a digital IC) to generate tuned perturbation $\vert$ Filter is used to further suppress any relaxation oscillation}}
\end{figure}

Fig.5 shows time dependent negative perturbation of around 7.5mA generated at the constant current node, which corresponds to node $\Delta$ in Fig.1. The negative perturbation at $\beta$is applied at the constant current node $\Delta$ by OPAMP and Logic block through a bias tee network.
\begin{figure}[ht!]
\includegraphics[width=9cm]{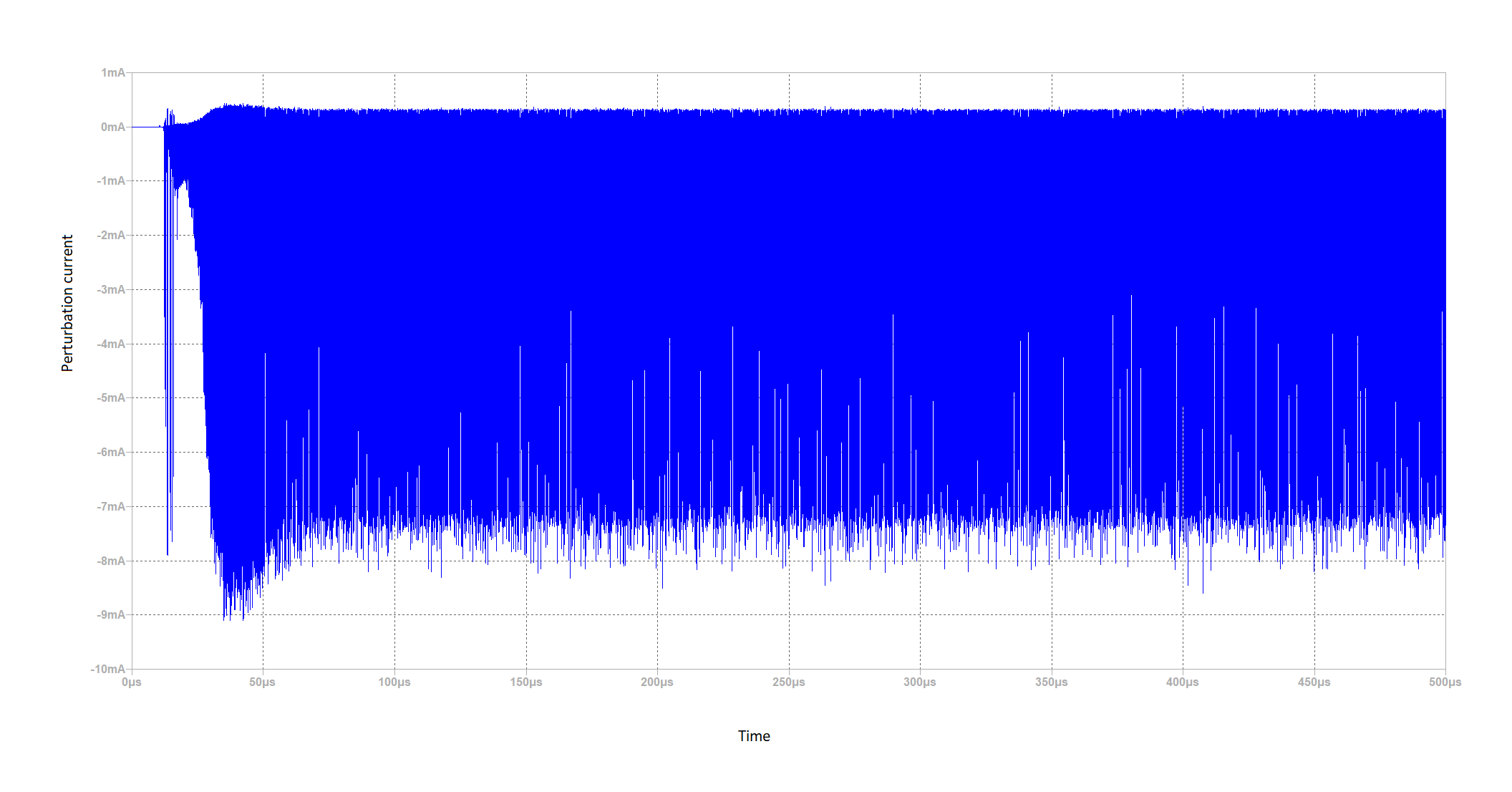}
\caption{\emph{Simulated negative perturbation applied at node $\beta \vert$ the plot shows multiple sub-nanosecond perturbation}}
\end{figure}
\begin{figure}[ht!]
\includegraphics[width=9cm]{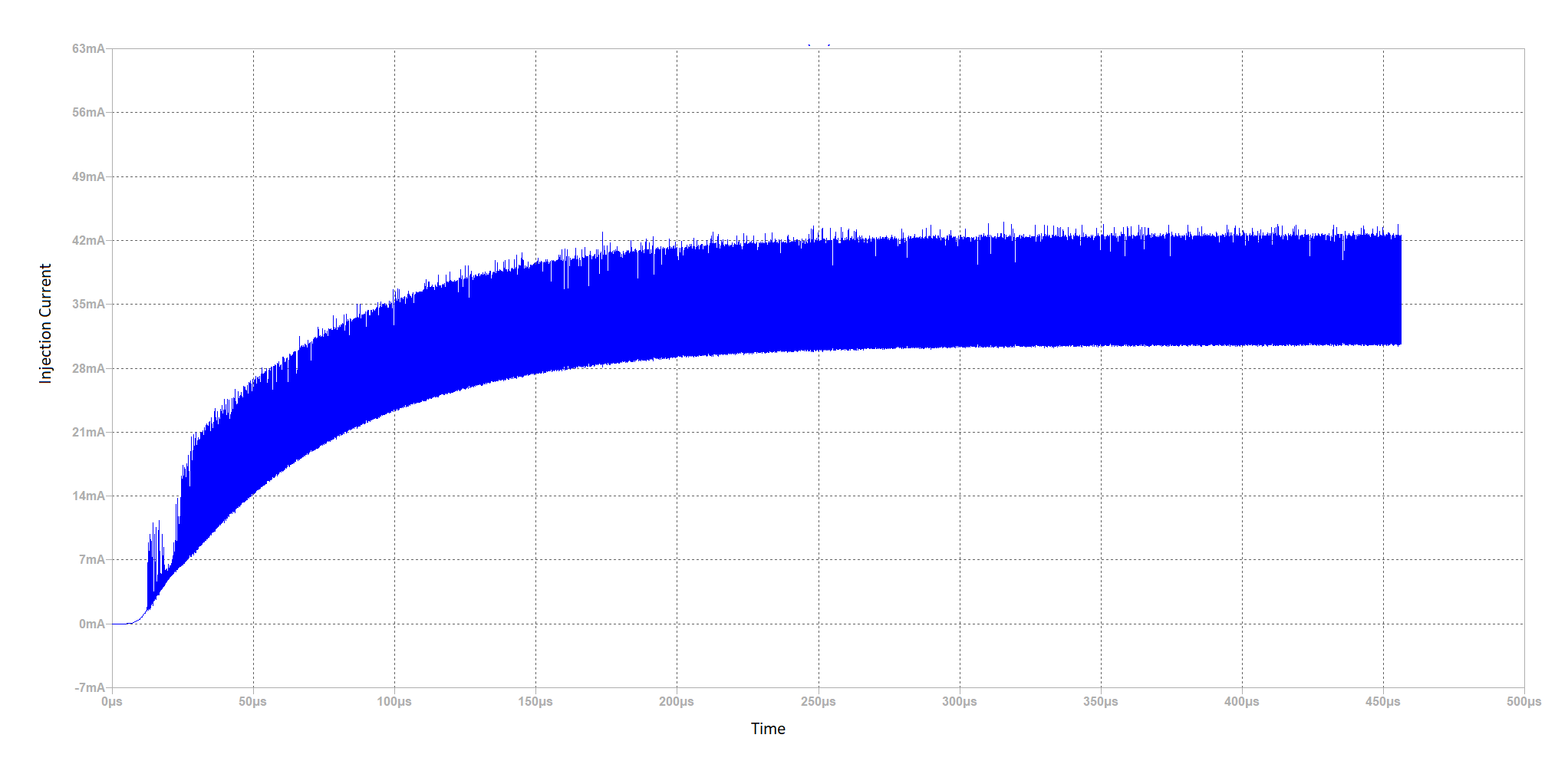}
\caption{\emph{Simulated current excitation profile at node $\psi \vert$ The plot shows sub-nanosecond perturbation on a constant excitation of 31mA}} 
\end{figure}
Fig.6 represents excitation current into the laser diode connected at the output node $\psi$ in Fig.1.The baseline current of 31mA represents constant current excitation from node $\alpha$.
Fig.7 represents zoomed in version of excitation shown in Fig.6. Sub-nanosecond pulse of FWHM 600ps can be observed. 
\begin{figure}[ht!]
\includegraphics[width=8cm]{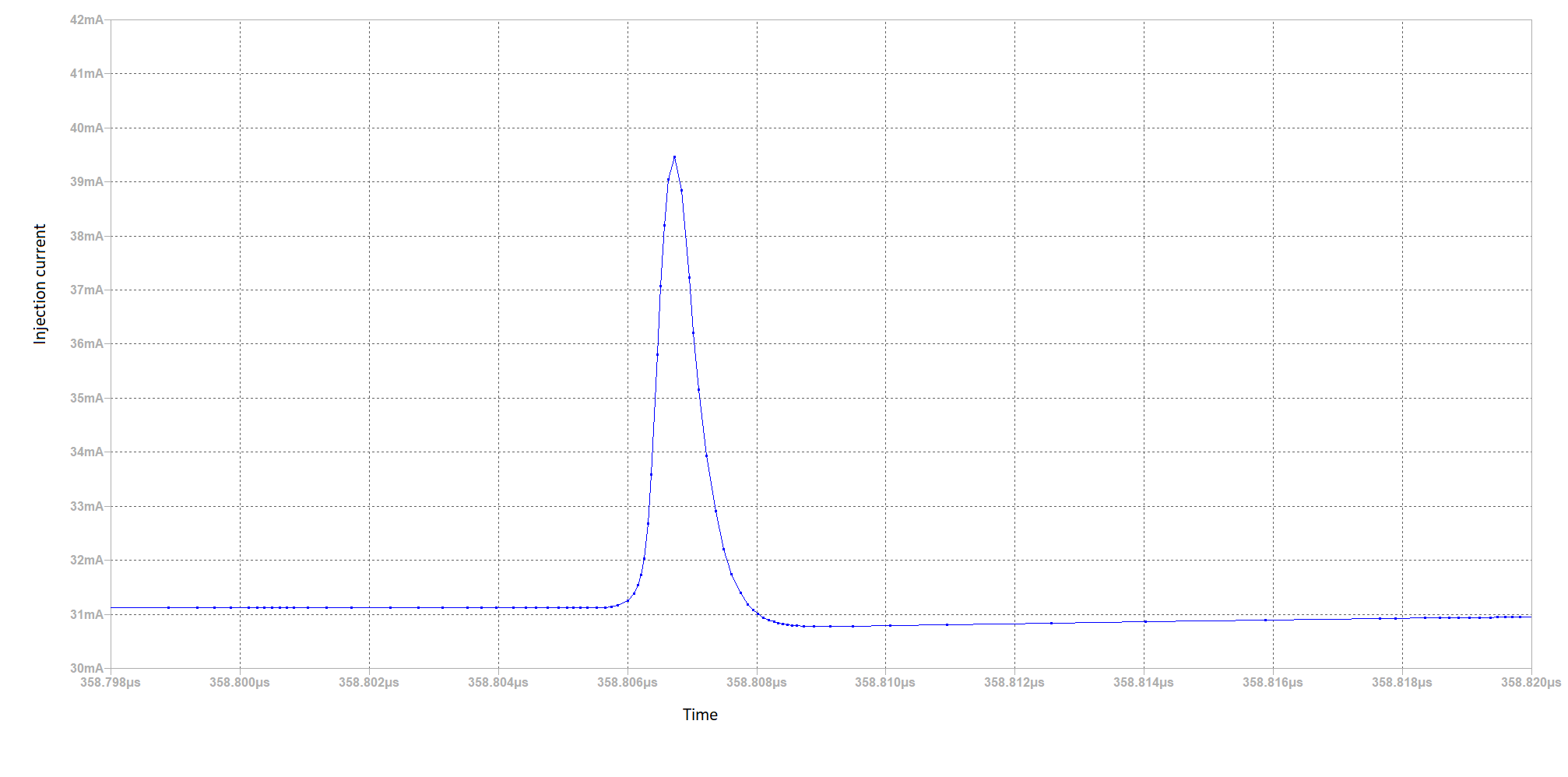}
\caption{ \emph{Simulated sub-nanosecond pulse excitation as seen on node $\psi$ $\vert$ The plot is zoomed in version of Fig.6}}
\end{figure}
Simulation model does not completely consider parasitic effect of the PCB as operating frequency is 100KHz. In addition, any deviation in IC parameters from its model considered for simulation will have impact on the observed pulse characteristics.
\begin{figure}[ht!]
\includegraphics[width=9cm]{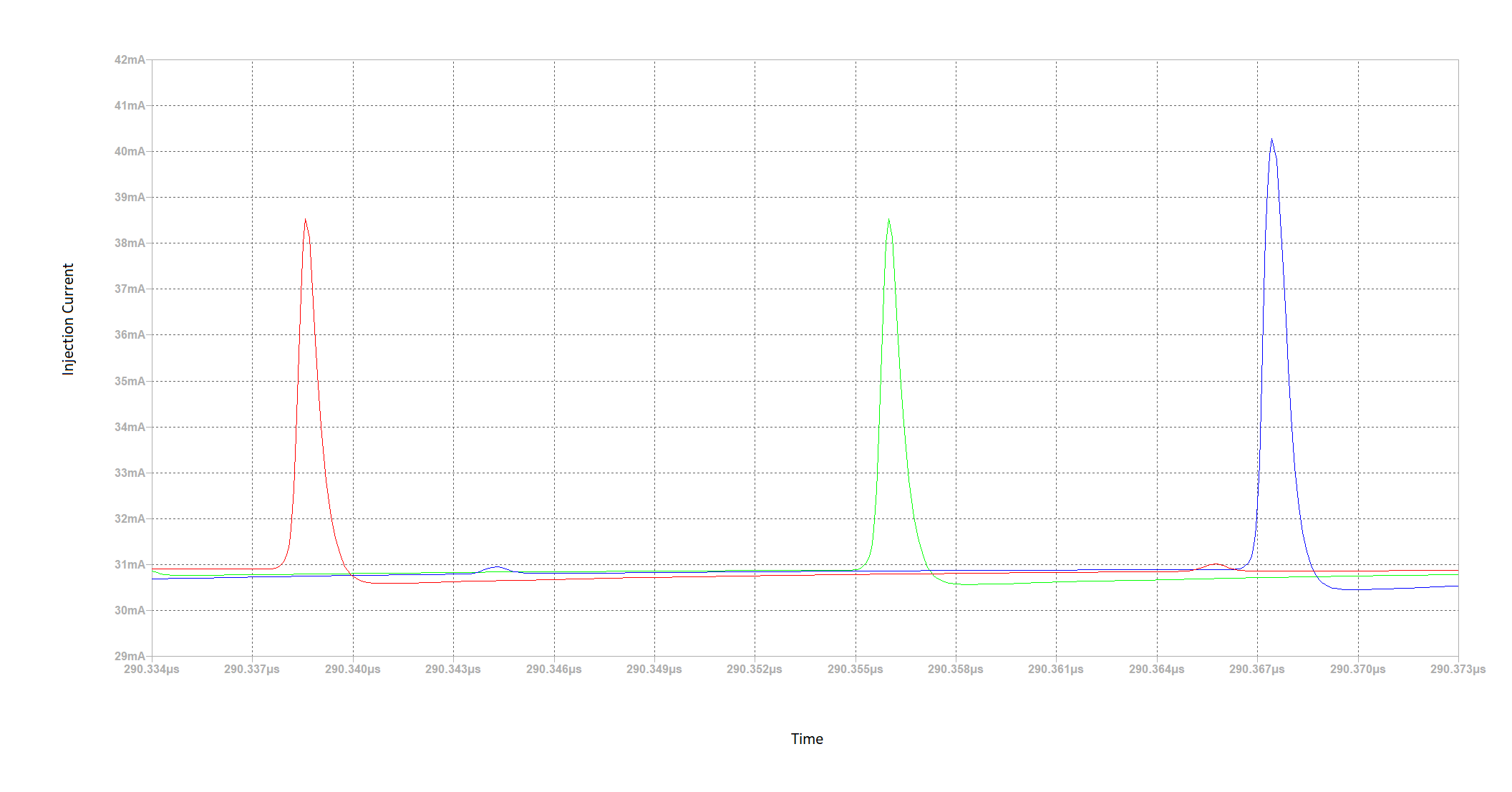}
\caption{\emph{Simulation of tun-ability in pulsed excitation $\vert$ One of the controlled parameters was varied for tuning the excitation pulse. Plot in red and green demonstrate tun-ability in delay and plot in blue represents tun-ability in amplitude of the perturbation. Amplitude tuning can also be achieved by tuning the constant current excitation, which represents shifting the base current}}
\end{figure}
Fig.8 represents tun-abilty achieved in the laser excitation. Excitation delay was tuned by approximately 8ns (red and blue). Moreover, amplitude of the pulse was tuned from 41.5mA to 39.2mA. Amplitude tuning can also be achieved by adjusting the constant excitation applied at node $\alpha$.

\section{Experiment and Results}

\begin{figure}[ht!]
\includegraphics[width=9cm]{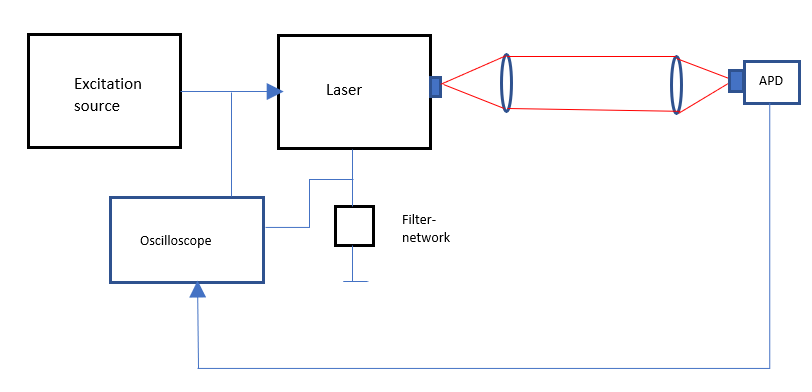}
\caption{\emph{Experimental setup for estimating current injection and output power of the laser diode $\vert$ Excitation source represents node $\psi$. Differential current into the laser diode and APD, which detects focused laser ( in red), is monitored using an oscilloscope}}
\end{figure}
Characterization of short-duration pulse has been discussed in [8].Experiment was carried out to verify sub-nanosecond, indistinguishable pulse generation from the laser diode, by measuring differential voltage across its terminals using differential probe D420-A-PB with a 4GHz DX20 tip connected to 4GHz Lecroy 640Zi Waverunner.Profile of emitted optical pulses was estimated using an active probe ZS2500 with 2GHz bandwidth and a custom built SAP500 APD detector operating in linear region. Since our receiver SAP500 APD is bandwidth limited, it does not provide accurate estimation of actual pulse width of the sub-nanosecond optical pulse. However, the profile can be verified from profile of the excitation perturbation across the laser diode. 
Fig.9 represents schematic of the experimental setup.Excitation source with pre-biased, sub-nanosecond current perturbation  drives HL6748MG, a 670nm laser diode (For cryptography application, it will be interesting to analyze coherence length and coherence time of this low cost diode). In addition, a filter is connected in series with the laser diode to sense current and for suppressing any oscillation. Also, feedback from the current sense can be used to track any change in temperature. For our application in automotive framework, we need extremely low cost implementation. Since the laser driver is driven near to the threshold region, and that the pulsed excitation is less than 1ns with frequency of 100KHz, average case temperature of laser diode will be near to  the room temperature. This way temperature control will not be needed.
Emission from the laser diode is monitored using custom made SAP500 based detector with the APD biased in linear region. The detector was selected based on efficiency corresponding to lasing wavelength of the laser diode. 
Fig.10 characterizes differential injection pulse through the laser diode. The shortest FWHM of the excitation pulse was observed to be 496ps. DC offset current due to pre-bias has been subtracted to underline pulse characteristics. 

\begin{figure}[ht!]
\includegraphics[width=9cm]{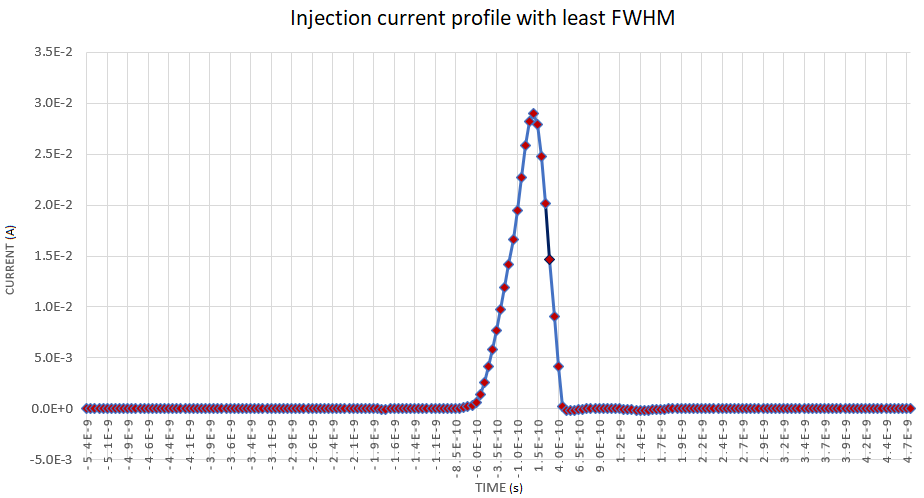}
\caption{\emph{Measured current excitation profile with least FWHM of 496ps $\vert$ constant current excitation has been subtracted to underline pulse characteristics}}
\end{figure}

Fig.11 demonstrates fine tuning of delay between current excitations by 60ps. Voltage level of 2.346V on zero reference of the excitation, was shifted to 60ps. Theoretically, the design should be able to attain any tuning required.
\begin{figure}[ht!]
\includegraphics[width=9cm]{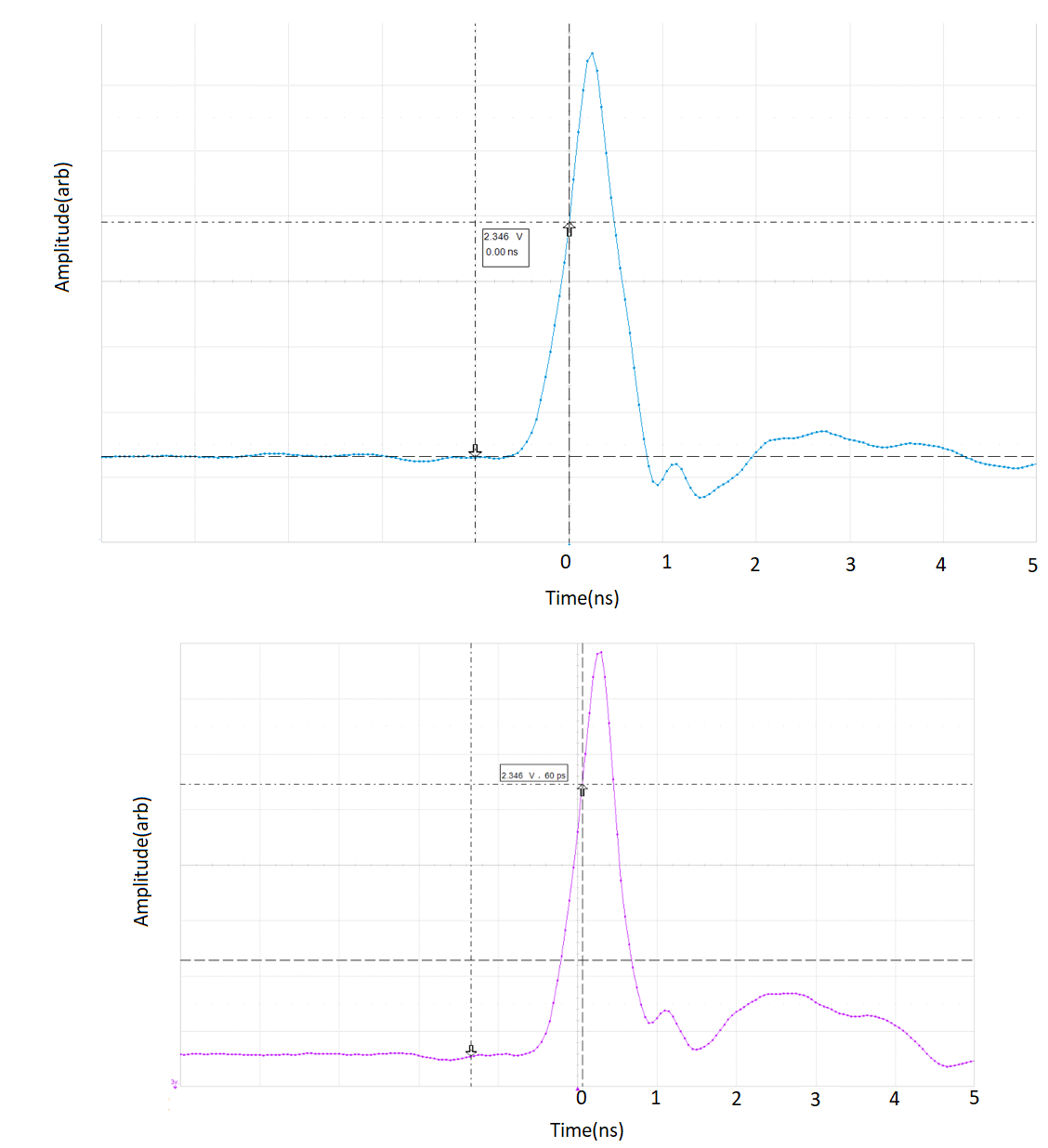}
\caption{\emph{Measurement of adjusted shift between two injection current profiles}}
\end{figure}

Fig. 12 represents typical response of excited laser diode on a APD, biased in linear region. Typical linewidth as measured on the APD is 525ps. This unexpectedly large line-width can be attributed to 500ps of rise and fall time of the APD itself. Side oscillation is due to parasitic impedance on the PCB. Similarly, Fig. 13 represents APD response of another indistinguishable excitation with higher amplitude.Two indistinguishable pulses need to have same characteristics in terms of shape, FWHM etc. .

\begin{figure}[ht!]
\includegraphics[width=9cm]{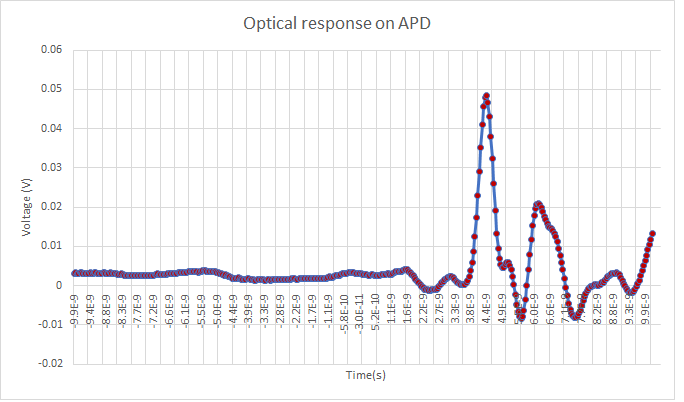}
\caption{\emph{Optical response of sub-nanosecond laser excitation as measured on APD ( rise time and fall time 500ps) $\vert$ Additional peaks are due to parasitic impedance on the board}}
\end{figure}

\begin{figure}[ht!]
\includegraphics[width=9cm]{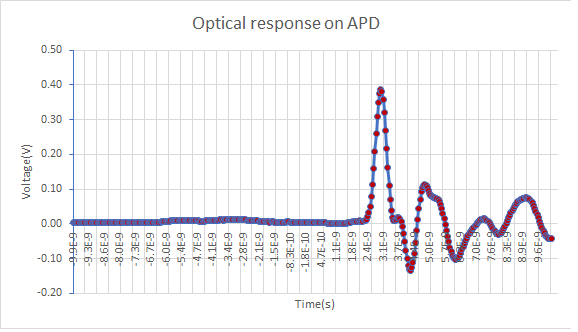}
\caption{\emph{Indistinguishable response measured on APD $\vert$ Additional peaks are due to parasitic impedance on the board}}
\end{figure}

Fig. 14 represents a comparative graph for amplitude normalized APD response. APD responses correspond to two different level of excitation into the laser diode. Graph corresponding to state2 has been shifted to overlap with state1, for better comparison.
\begin{figure}[ht!]
\includegraphics[width=9cm]{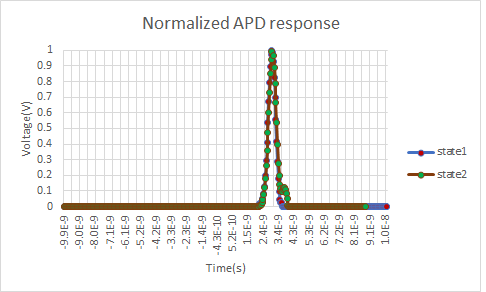}
\caption{\emph{Comparison of indistinguishability in temporal domain for two different laser ex citations$\vert$ APD responses were generated by varying the amplitude of excitation in the laser diode$\vert$ Amplitude of the graphs are normalized and shifted for better comparison. Further, oscillation due to parasitic inductance on the PCB has been removed}}
\end{figure}

Fig. 15 represents cumulative distribution function for state1 and state2 (fig. 14).For two-sample Kolmogorov-Smirnov test, with D as 0.028, p-value as 0.998, for $\alpha$=0.05. Since p-value is greater than $\alpha$, it an acceptable hypothesis that both states came from same distribution.

\begin{figure}[ht!]
\includegraphics[width=9cm]{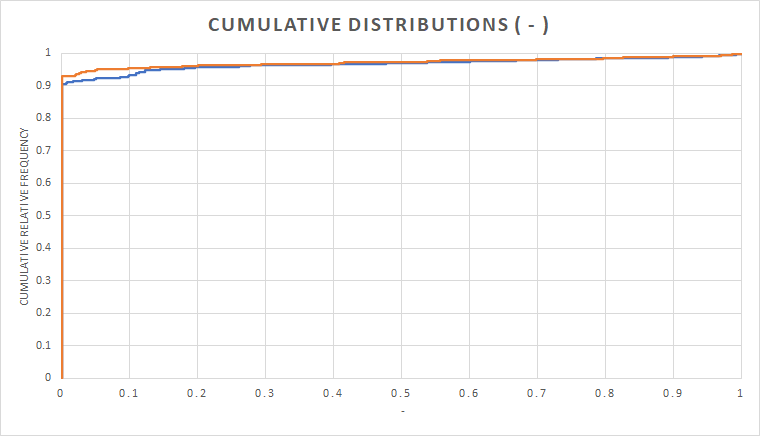}
\caption{\emph{Cumulative distribution function for state1 and state2 $\vert$ Two sample Kolmogorov-Smirnov maximum deviation estimated to be 0.0281, and p-value of 0.9979, for $\alpha$=0.05 $\vert$ since p-value is greater than $\alpha$ it is an acceptable hypothesis that state1 and state2 belongs to same distribution.}}
\end{figure}

 Since indistinguishability in terms of waveform shape is critical for our application, and so far we have considered indistinguishability in temporal domain. We have not considered any distinguishability in frequency domain due to chirps. It will be interesting to study impact on spectral indistinguishability using proposed approach.

\section{Conclusion}
We have designed a compact (4.5cmx3cm), sub-nanosecond, tun-able-indistinguishable laser pulse generator with excitation shortest FWHM measured around 496ps (Ideally, it will be less than 496ps if we consider contribution due to probe effect).Further, this can also be used to generate shape tun-able excitation for active quenching of APD in single photon detection experiment, and to perform photon number resolution using APD.Indistinguishability in spectral domain, using this technique, should be explored.

\section{Acknowledgement}
This study is funded by Robert Bosch LLC.

\section{References}
\begin{enumerate}
    \item Yin, HL., Zhou, MG., Gu, J. et al. Tight security bounds for decoy-state quantum key distribution. Sci Rep 10,
    14312 (2020). https://doi.org/10.1038/s41598-020-71107-6
    \item Wei, Z., Wang, W., Zhang, Z. et al. Decoy-state quantum key distribution with biased basis choice. Sci Rep 3, 2453 (2013). https://doi.org/10.1038/srep02453
    \item  Decoy State Quantum Key Distribution, Lo et al. 	arXiv:quant-ph/0411004\\
    \item Kardynał, B., Yuan, Z.  Shields, A. An avalanche‐photodiode-based photon-number-resolving detector. Nature Photon 2, 425–428 (2008). https://doi.org/10.1038/nphoton.2008.101
    \item Richard A. Linke, Alan H. Gnauck, "High Speed Laser Driving Circuit And Gigabit Modulation Of Injection Lasers," Proc. SPIE 0425, Single Mode Optical Fibers, (8 November 1983); doi: 10.1117/12.936224
    \item Smale, S.. “On the mathematical foundations of electrical circuit theory.” Journal of Differential Geometry 7 (1972): 193-210.
    \item J. Katz, S. Margalit, C. Harder, D. Wilt and A. Yariv, "The intrinsic electrical equivalent circuit of a laser diode," in IEEE Journal of Quantum Electronics, vol. 17, no. 1, pp. 4-7, January 1981, doi: 10.1109/JQE.1981.1070628.
    \item Signal sources, conditioners and power circuitry, Jim Williams, November 2004
    \item Joachin Wabnig et al. 2015 Secured Wireless Communication US 9,641,326,B2,(\url{https://patft.uspto.gov/netacgi/nph-Parser?Sect1=PTO2&Sect2=HITOFF&p=1&u=%2Fnetahtml%2FPTO%2Fsearch-bool.html&r=1&f=G&l=50&co1=AND&d=PTXT&s1=wabnig.AANM.&OS=AANM/wabnig&RS=AANM/wabnig})
    \item Bunandar, Darius et al.2016 Apparatus and methods for quantum key distribution, US10,158,481 B2, (\url{https://patft.uspto.gov/netacgi/nph-Parser?Sect1=PTO2&Sect2=HITOFF&p=1&u=%2Fnetahtml%2FPTO%2Fsearch-adv.htm&r=25&f=G&l=50&d=PTXT&S1=bunandar&OS=bunandar&RS=bunandar})
    \item Nordholt et al. 2017, Quantum Communication system with integrated photonic devices. US 9, 819,418 B2, (\url{https://worldwide.espacenet.com/publicationDetails/originalDocument?FT=D&date=20171114&DB=&locale=en_EP&CC=US&NR=9819418B2&KC=B2&ND=4#})
    \item Yuan et al. 2017, Interference system and an interference method, US 9,696, 133 B2 (\url{https://worldwide.espacenet.com/publicationDetails/originalDocument?FT=D&date=20170704&DB=&locale=en_EP&CC=US&NR=9696133B2&KC=B2&ND=4#})

\end{enumerate}
\end{document}